# ENVIRONMENTAL CONTOURS BASED ON KERNEL DENSITY ESTIMATION

Andreas F. Haselsteiner[1,2], Jan-Hendrik Ohlendorf[1,2], Klaus-Dieter Thoben[1,2]
[1] University of Bremen, BIK - Institute for Integrated Product Development, 28359 Bremen, Germany,
a.haselsteiner@uni-bremen.de, johlendorf@uni-bremen.de, tho@biba.uni-bremen.de
[2] ForWind – Center for Wind Energy Research of the Universities of Oldenburg, Hannover and Bremen

**Summary**

An offshore wind turbine needs to withstand the environmental loads, which can be expected during its life time. Consequently, designers must define loads based on extreme environmental conditions to verify structural integrity. The environmental contour method is an approach to systematically derive these extreme environmental design conditions. The method needs a probability density function as its input. Here we propose the use of constant bandwidth kernel density estimation to derive the joint probability density function of significant wave height and wind speed. We compare kernel density estimation with the currently recommended conditional modeling approach. In comparison, kernel density estimation seems better suited to describe the statistics of environmental conditions of simultaneously high significant wave height and wind speed. Consequently, an environmental contour based on kernel density estimation does include these environmental conditions while an environmental contour based on the conditional modeling approach does not. Since these environmental conditions often lead to the highest structural responses, it is especially important that the used method outputs these conditions as design requirements.

## 1. Introduction

An offshore wind turbine needs to withstand the loads that the marine environment exerts on it. In the design phase of such a turbine, designers have to fulfill standards, which require the turbine to withstand the extreme environmental conditions that can be expected to occur with a return period of 50 years [7]. The environmental contour method is an approach, which helps designers to select such extreme environmental conditions based on a given data set.

There exists a whole family of different specific environmental contour methods since different definitions for the exceedance region can be used [4]. Specific environmental contour methods are for example the *inverse first-order reliability method* (IFORM) [11], the constant probability density approach [2], a Monte Carlo simulation based hyper-planes approach [6] or the *highest density contour* (HDC) method [4]. The first step before applying a specific environmental contour technique, however, is always to estimate the probability density function from a measured or simulated data set. Traditionally, the probability density function is estimated by following the so-called *conditional modeling approach* (CMA) [1,2] meaning that starting from one independent random variable the other variables are defined to be conditional on other variables. Last year, however, Eckert-Gallup and Martin [3] proposed to use *kernel density estimation* (KDE) with adaptive bandwidth selection as the basis for environmental contours. This non-parametric density estimation approach allows for more flexibility since it does not define any dependencies *apriori*.

Motivated by the promising results Eckert-Gallup and Martin [3] achieved for wave data of a buoy located offshore of Northern California, we explore if KDE is also well suited for the combination of wind and wave data. Further, we will use the simpler constant bandwidth KDE and compare our results to the established *conditional modeling approach*.

## 2. Data and Methods

In summary the approach we present here consists of three steps: (i) First, we estimate the joint probability density function based on hindcast data using kernel density estimation. (ii) Then we compute the environmental contour by applying the highest density contour method. (iii) Finally, we select a finite number of extreme environmental design conditions along the contour's path.

2.1 Hindcast data
We analyze the environmental conditions at two locations in the Southern North Sea, at the research platform FINO3 and at the wind farm Trianel Windpark Borkum (short "Trianel"', Fig. 1*a*). The basis for the analysis are 49 years of data from the openly available coastDat-1 hindcast [5,10]. We use a continuous time series, which starts on January 1st 1958 and ends on December 31st 2006. Among the available variables we select significant wave height, $H_s$, and wind speed, $V$, (at a height of 10 m above the sea level, averaged over 1 hour, $n = 429{,}528$, Fig. 1*b*). These variables are especially important to define design loads and are therefore dealt with extensively in standards and guidelines [2,7].

2.2 Kernel density estimation
Kernel density estimation is a non-parametric method to estimate the probability density function based on a given data set. While KDE was first developed as a univariate method, it was later generalized to multivariate statistics and is now a well-established method for multivariate density estimation [8]. Here we use bivariate kernel density estimation with Gaussian kernels with a constant bandwidth for each dimension *i* (Fig. 1*c*). The bandwidth, $b_i$, is chosen to be two times Silvermans's rule of thumb [9], which is based on the number of data points, $n$, and the standard deviation, $\sigma_i$:

$$b_1 = 2\sigma_1 \left(\frac{1}{n}\right)^{\frac{1}{6}}$$
$$b_2 = 2\sigma_2 \left(\frac{1}{n}\right)^{\frac{1}{6}}$$

We use Matlab's (version R2017b Prerelease, The MathWorks, USA) *ksdensity* function to perform the computation and choose a grid resolution of 0.1 m × 0.1 m s$^{-1}$. For comparison, we also estimate the probability density using the conditional modeling approach. Consequently, we use the $H_s$-$V$-density function that is recommended by the certifying organization DNVGL [2] and has been proposed by Bitner-Gregersen [1].

2.3 Environmental contour
We use the highest density contour method [4] to derive environmental contours with return periods of 1, 50 and 500 years. The input for the environmental contour computation is the probability density function derived by either using kernel density estimation or the conditional modeling approach. An environmental contour's return period describes at which recurring time period one can expect the occurrence of an environmental condition outside the environmental design region (Fig. 2a). A highest density contour is defined to have constant probability density along its path and to enclose a region of exactly 1 - $\alpha$ probability, with $\alpha$ being the exceedance probability [4] (Fig. 2b).

2.4 Design conditions
As the last step, we select 9 extreme environmental design conditions per environmental contour. We do this by introducing a polar coordinate system which has its origin, $O$, at the median of the raw data, $O$ = (median($H_s$), median($V$)). Then we find the extreme environmental design conditions by defining normalized polar angles, $\varphi^* = \{0,...,90°\}$, and following each line of angle $\varphi^*$ until it intersects with the environmental contour. Additionally, we search along each contour for the condition of maximum significant wave height, max($H_s$), and for the condition of maximum wind speed, max($V$).

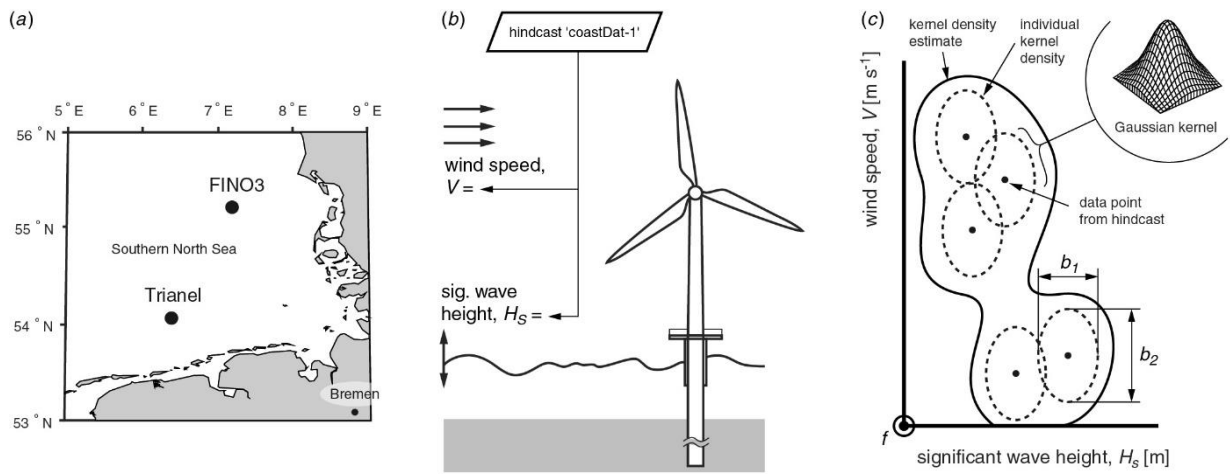

Fig. 1. **Used data.** (*a*) Two locations in the North Sea are analyzed, FINO3 and Trianel. (*b*) The environmental variables wind speed, *V*, and significant wave height, $H_s$, are taken from the coastDat-1 hindcast data set [5]. (*c*) Based on the hindcast data, probability density, *f*, is estimated using bivariate kernel density estimation with Gaussian kernels of bandwidth *b*.

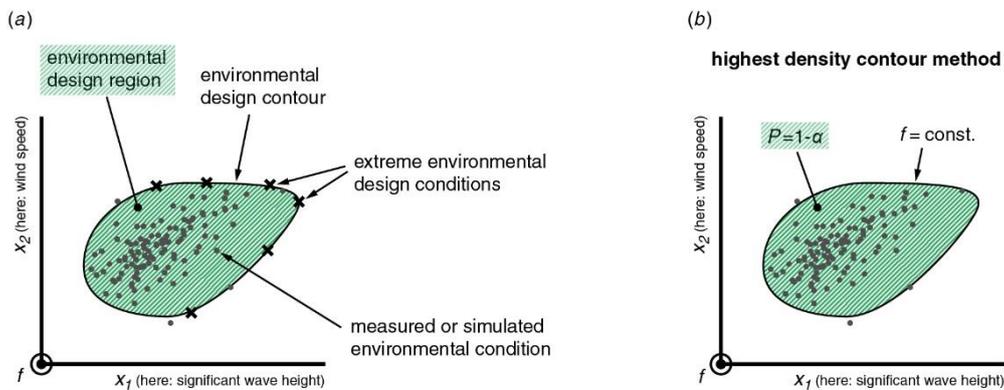

Fig. 2. **Environmental contour method.** (*a*) General concept of an environmental contour and the corresponding terms. (*b*) Specific definition of the highest density contour method. Adapted from [4].

## 3. Results

Visual impression suggests that the derived contours based on KDE describe the statistics of the hindcast data well (Fig. 3*a*). At FINO3 the 1-year contour excludes 21 data points and the 50-year contour includes all data points. At Trianel 22 data points are outside the 1-year contour and all data points are included by the 50-year contour. On the other hand, at both analyzed locations the 50-year contour based on the CMA does not include hundreds of data points (FINO3: 520, Trianel: 273; Fig. 3*b*). If a 49-year random data set were derived from a given probability distribution it would be extremely unlikely that more than 10 data points exceed a 50-year contour (1 - $F_{Binomial}$($n$ = 429,528, $p$ = 1 / (50×365.25 ×24, $k$ = 10) ≈ 8×10$^{-9}$). Consequently, it can be questioned whether the probability density function derived from the CMA is a good approximation of the true probability density function.

While there are data points outside the contour based on the CMA at a variety of locations, the effect appears to be most severe for data points of simultaneously high significant wave height, $H_s$, and wind speed, $V$. The contour based on KDE, on the other hand, does include these data points.

The maximum values along the contour differ between 12 and 18 % when the 50-year contours based on KDE and the CMA are compared. In the case of FINO3 the maximum value for the significant wave height, $H_s$, is 10.35 m for the contour based on the CMA and 12.15 m for the contour based on KDE (Tab. 1). In the case of Trianel the values are 9.75 m and 10.95 m respectively. The maximum wind speed along the contour is higher for the contour based on KDE as well (26.15 m s$^{-1}$ vs 30.25 m s$^{-1}$ for FINO3 and 26.85 m s$^{-1}$ vs 30.35 m s$^{-1}$ for Trianel).

| $\varphi^*$ | CMA $H_s$ [m] | V [m s$^{-1}$] | KDE $H_s$ [m] | V [m s$^{-1}$] |
|---|---|---|---|---|
| FINO3: | | | | |
| 0 | 6.15 | 8.05 | 6.65 | 8.05 |
| 15 | 7.25 | 11.95 | 7.05 | 11.75 |
| 30 | 9.05 | 18.85 | 12.05 | 23.05 |
| 45 | 8.65 | 25.95 | 10.35 | 29.95 |
| 60 | 5.05 | 23.55 | 6.05 | 27.95 |
| 75 | 2.75 | 20.55 | 2.85 | 21.85 |
| 90 | 1.35 | 15.95 | 1.35 | 17.75 |
| max($H_s$) | **10.35** | 25.25 | **12.15** | 24.75 |
| max($V$) | 9.85 | **26.15** | 10.65 | **30.25** |
| Trianel: | | | | |
| 0 | 5.85 | 7.65 | 4.95 | 7.65 |
| 15 | 7.25 | 12.05 | 6.35 | 11.35 |
| 30 | 9.05 | 19.95 | 10.95 | 22.95 |
| 45 | 8.15 | 26.65 | 9.45 | 30.05 |
| 60 | 4.45 | 22.75 | 5.35 | 27.15 |
| 75 | 2.35 | 18.95 | 2.55 | 21.15 |
| 90 | 1.25 | 15.65 | 1.25 | 17.15 |
| max($H_s$) | **9.75** | 24.95 | **10.95** | 26.15 |
| max($V$) | 8.85 | **26.85** | 9.35 | **30.35** |

Tab. 1. Comparison of the extreme environmental design conditions derived by following the conditional modeling approach (CMA) versus kernel density estimation (KDE).

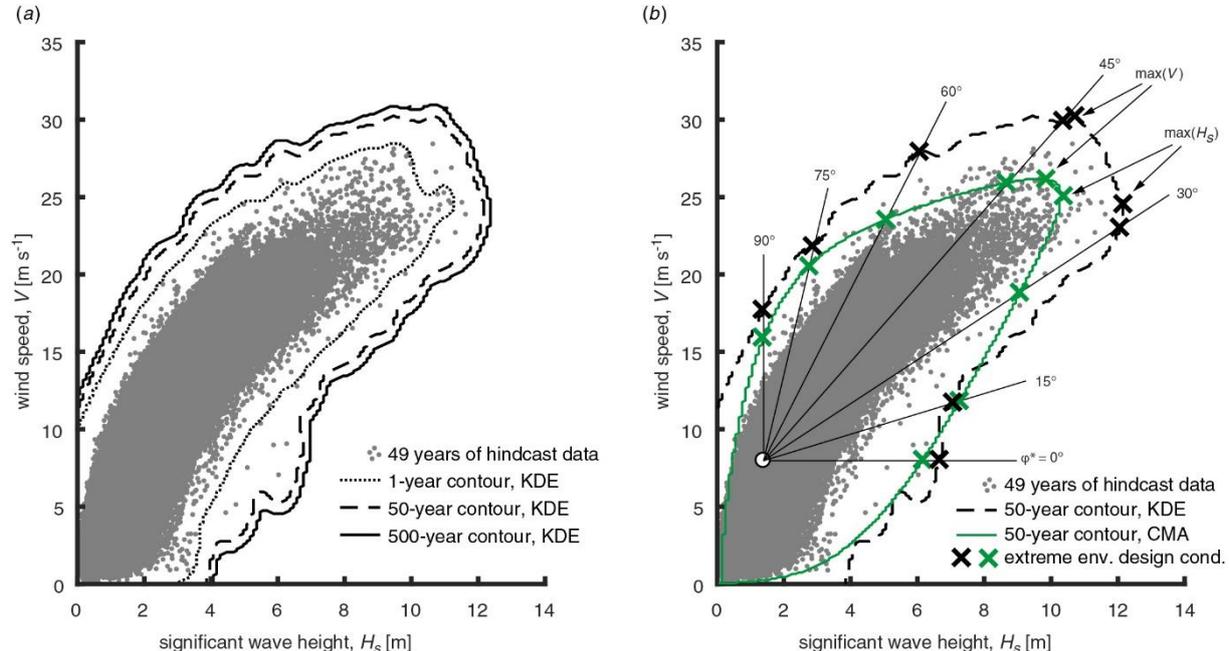

Fig. 3. **Environmental contours for the location FINO3.** (*a*) Contours of increasing return periods, which are based on kernel density estimation. (*b*) Comparison between a 50-year contour based on kernel density estimation (KDE) and a contour based on the conditional modeling approach (CMA).

## 4. Discussion

Environmental conditions of simultaneously high significant wave height and wind speed often lead to the highest structural responses. Therefore, it is especially important that the chosen probability density function represents the measured high-wind-speed-high-significant-wave-height data well. Visual impression as well as the probability of data points exceeding a 50-year contour suggests that kernel density estimation is advantageous compared to the conditional modeling approach. Going further, the more representative density function based on KDE leads to a more representative environmental contour, which leads to more representative extreme environmental design conditions. These extreme environmental design conditions are part of the requirements of an offshore wind turbine as they are the basis for the design load cases, which have to be considered when structural integrity is analyzed [7]. Consequently, the choice of extreme environmental design conditions ultimately influences the resulting wind turbine design and high quality extreme environmental design conditions enable high quality engineering design.

Eckert-Gallup and Martin [3] have used adaptive kernel density estimation since they perceived areas of sparse data to be especially complicated to handle due to the sensitivity of kernel density estimation to the chosen bandwidth. Using Abramson's estimator their method chooses wider bandwidths at areas of sparse data. While we do not make a direct comparison to adaptive KDE here, we show that the simpler constant bandwidth KDE also achieves satisfying results. Here we have chosen a relative wide bandwidth (2×Silverman's rule of thumb). Consequently, we might oversmooth the probability density function at areas of high data density. Adaptive KDE would use a smaller bandwidth at these areas. However, since the environmental contour method deals with extreme environmental conditions, which lie in areas of low data density, we do not perceive oversmoothing in the areas of high data density as a problem.

## 5. Conclusions

Constant bandwidth KDE can be used to estimate the probability density of environmental conditions, which are relevant for structural design. In comparison to the conditional modeling approach, which is recommended in current guidelines, KDE seems better suited to handle environmental conditions of simultaneously high significant wave height and wind speed. Since these environmental conditions often lead to the highest structural responses, it is especially important that the probability density function and the resulting environmental contour accurately cover them.

## References

[1] Bitner-Gregersen E.M. (1991): Joint environmental model for reliability calculations. In: Proceedings of the International Offshore and Polar Engineering Conference, pp. 246-253.

[2] Det Norske Veritas (2010): Recommended practice – DNV-RP-C205 Environmental conditions and environmental loads. Tech. Rep.

[3] Eckert-Gallup A., Martin N. (2016): Kernel density estimation (KDE) with adaptive bandwidth selection of extreme sea states. In: OCEANS 2016 MTS/IEEE Monterey. Monterey, CA, USA, pp 1-5.

[4] Haselsteiner A.F., Ohlendorf J.-H., Wosniok W., Thoben K.-D. (2017): Deriving environmental contours from highest density regions. *Coastal Engineering* 123, 42-51.

[5] Helmholtz-Zentrum Geesthacht, Zentrum für Material- und Küstenforschung GmbH (2012): coastDat-1 Waves North Sea wave spectra hindcast (1948-2007). World Data Center for Climate (WDCC) at DKRZ.

[6] Huseby A.B., Vanem E., Natvig B. (2013): A new approach to environmental contours for ocean engineering applications based on direct Monte Carlo simulations. *Ocean Engineering* 60, 124-135.

[7] International Electrotechnical Commission (2009): Wind turbines – part 3: design requirements for offshore wind turbines. Tech. Rep. IEC 61400-3:2009-02.

[8] Scott D.W. (2015): Multivariate density estimation: theory, practice and visualization, 2nd Edition. Wiley, Hoboken, NJ, USA.

[9] Silverman B.W. (1998): Density estimation for statistics and data analysis. CRC press, London, UK, pp. 86-87.

[10] Weisse R. (2007): Wave climate and long-term changes for the Southern North Sea obtained from a high-resolution hindcast 1958-2002. *Ocean Dynamics* 57 (3), 161-172.

[11] Winterstein S.R., Ude T.C., Cornell C.A., Bjerager P., Haver S. (1993): Environmental parameters for extreme response: inverse FORM with omission factors. In: Proceedings, ICOSSAR-93. Innsbruck, Austria.

**Acknowledgements**

We thank O. Arend, D. Bode and M. Brink for reading earlier versions of this manuscript and giving critical feedback and W. Wosniok for fruitful discussions.